\documentclass[]{spie}  

\usepackage{amsmath,amsfonts,amssymb}
\usepackage{graphicx}

\newcommand {\micron}{\mbox{$\mu$m}}
\newcommand\arcsec{$^{\prime\prime}$}

\title{Performance of the extreme-AO instrument VLT/SPHERE and dependence on the atmospheric conditions}

\author[a]{J. Milli}
\author[b]{D. Mouillet}
\author[c,d]{T. Fusco}
\author[a]{J. H. Girard}
\author[e]{E. Masciadri}
\author[a]{E. Pena}
\author[c,d]{J-F. Sauvage}
\author[a]{C. Reyes}
\author[c]{K. Dohlen}
\author[b]{J-L. Beuzit}
\author[f]{M. Kasper}
\author[a]{M. Sarazin}
\author[g]{F. Cantalloube}
\author[h]{the SHARDDS team} 

\affil[a]{European Southern Observatory (ESO), Alonso de C\'ordova 3107, Vitacura, Casilla 19001, Santiago, Chile}
\affil[b]{Universit\'e Grenoble Alpes, CNRS, IPAG, F-38000 Grenoble, France}
\affil[c]{Aix Marseille Universit\'e, CNRS, LAM (Laboratoire d'Astrophysique de Marseille) UMR 7326, 13388, Marseille, France}
\affil[d]{ONERA, The French Aerospace Lab, BP72, 29 avenue de la Division Leclerc, 92322 Chatillon Cedex, France}
\affil[e]{Osservatorio Astrofisico di Arcetri, INAF, Largo Enrico Fermi 5, I-50125, Florence, Italy}
\affil[f]{ESO, Karl-Schwarzschild-Stra{\ss}e 2, 85748 Garching, Germany}
\affil[g]{Max-Planck-Institut fur Astronomie, Konigstuhl 17, 69117 Heidelberg, Germany}
\affil[h]{SPHERE High Angular Resolution Debris Disc Survey}

\authorinfo{Further author information, send correspondence to Julien Milli: jmilli@eso.org}

\begin{document} 
\maketitle

\begin{abstract}
SPHERE is the high-contrast exoplanet imager and spectrograph installed at the Unit Telescope 3 of the Very Large Telescope. After more than two years of regular operations, we analyse statistically the performance of the adaptive optics system and its dependence on the atmospheric conditions above the Paranal observatory, as measured by the suite of dedicated instruments which are part of the Astronomical Site Monitor and as estimated by the SPHERE real-time calculator. We also explain how this information can be used to schedule the observations in order to yield the best data quality and to guide the astronomer when processing his/her data. 
\end{abstract}

\keywords{SPHERE, extreme Adaptive Optics, Strehl, contrast, atmospheric conditions, AO4ELT Proceedings}

\section{INTRODUCTION}
\label{sec_intro}  

The Spectro-Polarimeter High-contrast Exoplanet REsearch \cite{Beuzit2008} (SPHERE) is  a high-contrast instrument dedicated to planet searches. It is fed with an extreme adaptive optics (AO) system called SAXO (Sphere Ao for eXoplanet Observation). It operates at a frequency up to 1.38\,kHz\footnote{The loop maximum frequency initially set to 1.2\,kHz at the start of operations was raised to 1.38\,kHz in November 2015 to improve the performance on bright targets since no hardware limitation was encountered} on bright targets with a 40x40 spatially filtered Shack-Hartmann wavefront sensor (WFS) and a 41x41 piezoelectric high-order deformable mirror. It can deliver a very high Strehl, above 90\%, and can correct perturbations induced by the atmospheric turbulence and from the internal aberrations of the instrument itself. SAXO feeds three subsystems within SPHERE: the Infra- Red Dual- beam Imager and Spectrograph (IRDIS\cite{Dohlen2008}), the Integral Field Spectrograph (IFS\cite{Claudi2008}) and the rapid-switching Zurich IMaging POLarimeter (ZIMPOL\cite{Thalmann2008}). A comprehensive description of the SAXO design can be found in Fusco et al. 2006\cite{Fusco2006}. On-sky performance, as measured during the commissionings, are also available in Fusco et al. 2016\cite{Fusco2016}. For this study, the goal is to present the overall statistical results gathered during the first two and a half years of regular operations at the Paranal Observatory. In section \ref{sec_scope}, we define the scope of this study, then we present the results in terms of Strehl in section \ref{sec_strehl} and contrast in section \ref{sec_contrast} before concluding in section \ref{sec_conclusions}.

\section{Method and sample definition}
\label{sec_scope}  

We analysed AO telemetry data from observations taken between January 1st 2015 and May 1st 2017 over almost two and a half years. For the sake of consistency, we restricted our analysis to observations made in the near-infrared with the IFS and/or the IRDIS subsystems. In such a case, in the visible arm of SPHERE, all the light is sent to the WFS with a mirror. Conversely, the optical subsystem ZIMPOL makes use of a grey beamsplitter or a $H_\alpha$ dichroic, which leads to a decrease in the WFS flux of 1.7 or 0.4 magnitude respectively. These near-infrared observations correspond to 200 000 telemetry data points, with each point corresponding to an average of 20 seconds. These data points are spread over 465 different nights and more than a thousand different stars. These data are also publicly available (with the exception of observations still under proprietary time or belonging to the Guarantied Time Observation of the SPHERE consortium) and can be downloaded from the SPHERE archive query form \href{http://archive.eso.org/wdb/wdb/eso/sphere/form}{http://archive.eso.org/wdb/wdb/eso/sphere/form} by entering \texttt{OBJECT,AO} in the user-defined input field \textup{DPR TYPE}. They consist of multi-extension fits files, with the data discussed in this analysis being contained in a binary table with the fits extension \texttt{AtmPerfParams}.

The AO telemetry data includes estimates from the real-time computer (hereafter RTC) which highlights different quantities such as: the Strehl ratio (in short Strehl) and additional atmospheric parameters including the seeing and the coherence time\footnote{The binary table contains the Fried parameter $r_0$ and the equivalent velocity $v$, from which the seing $\epsilon$ and the coherence time  $\tau_0$ can be derived using the formula $\epsilon=\frac{\lambda}{r_0}$ with $\lambda=500$nm and $\tau_0=0.31\frac{r_0}{v}$ from Roddier (1981)\cite{Roddier1981} .} which are discussed in this paper. Unless otherwise specified, the Strehl is defined in the H band at 1.6\micron, while the seeing and coherence time are defined at 500nm. These data points do not include the full telemetry data, including for instance the WFS slopes at each time stamp, which are saved only on a case by case basis in a recording up to two minutes long, due to the volume of files generated.
To be able to connect the AO performance with the brightness of the target, we also queried the V and R magnitude of the AO target star from the Simbad database\cite{Wenger2000}, as a proxy for the number of photons received by the WFS. The WFS sensitivity peaks in the R band, but we mostly used the V band magnitude in the rest of this analysis because the R magnitude is not available for all targets.

\begin{figure} [ht]
   \begin{center}
   \begin{tabular}{c c} 
   \includegraphics[height=6cm]{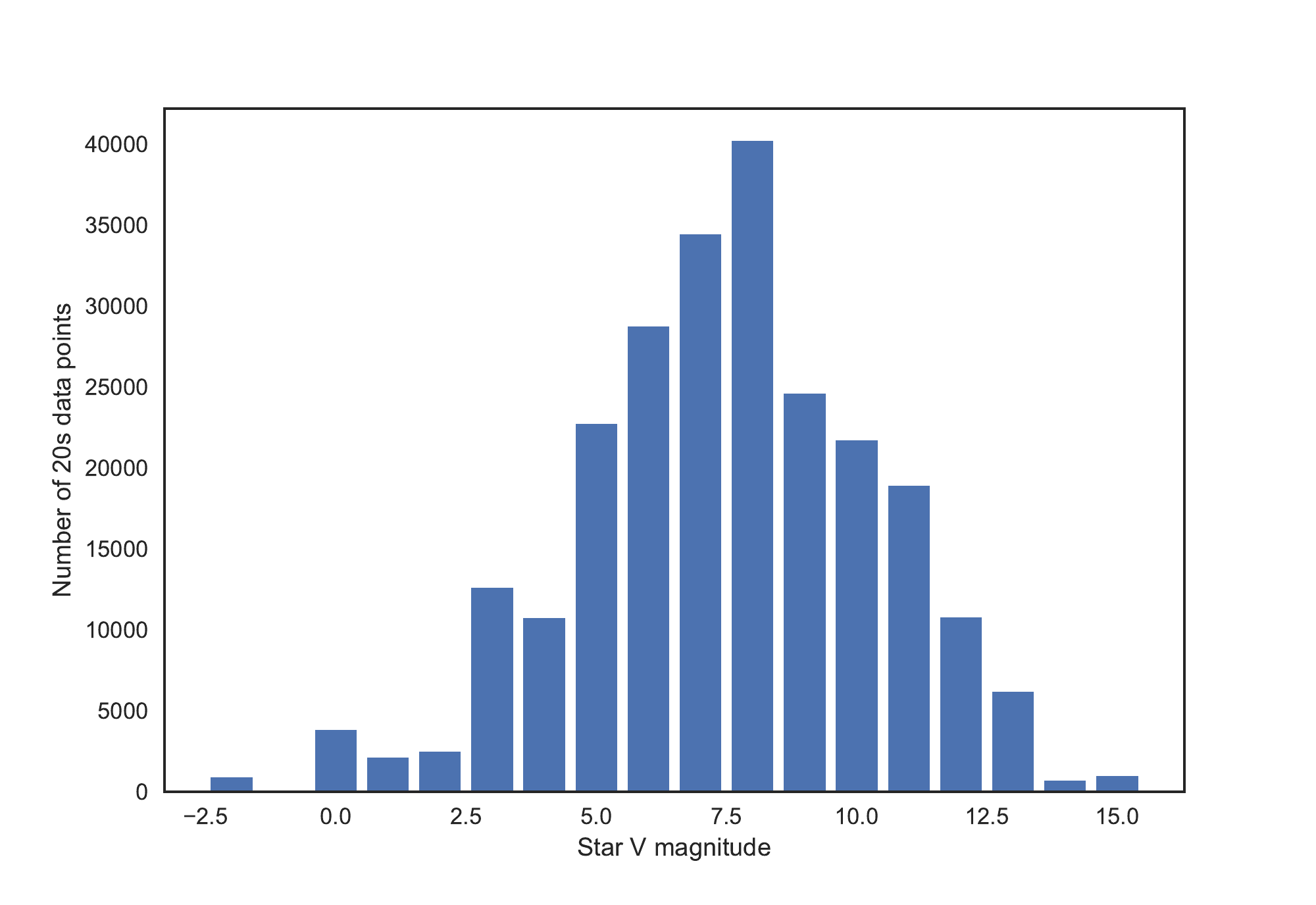} 
   \includegraphics[height=6cm]{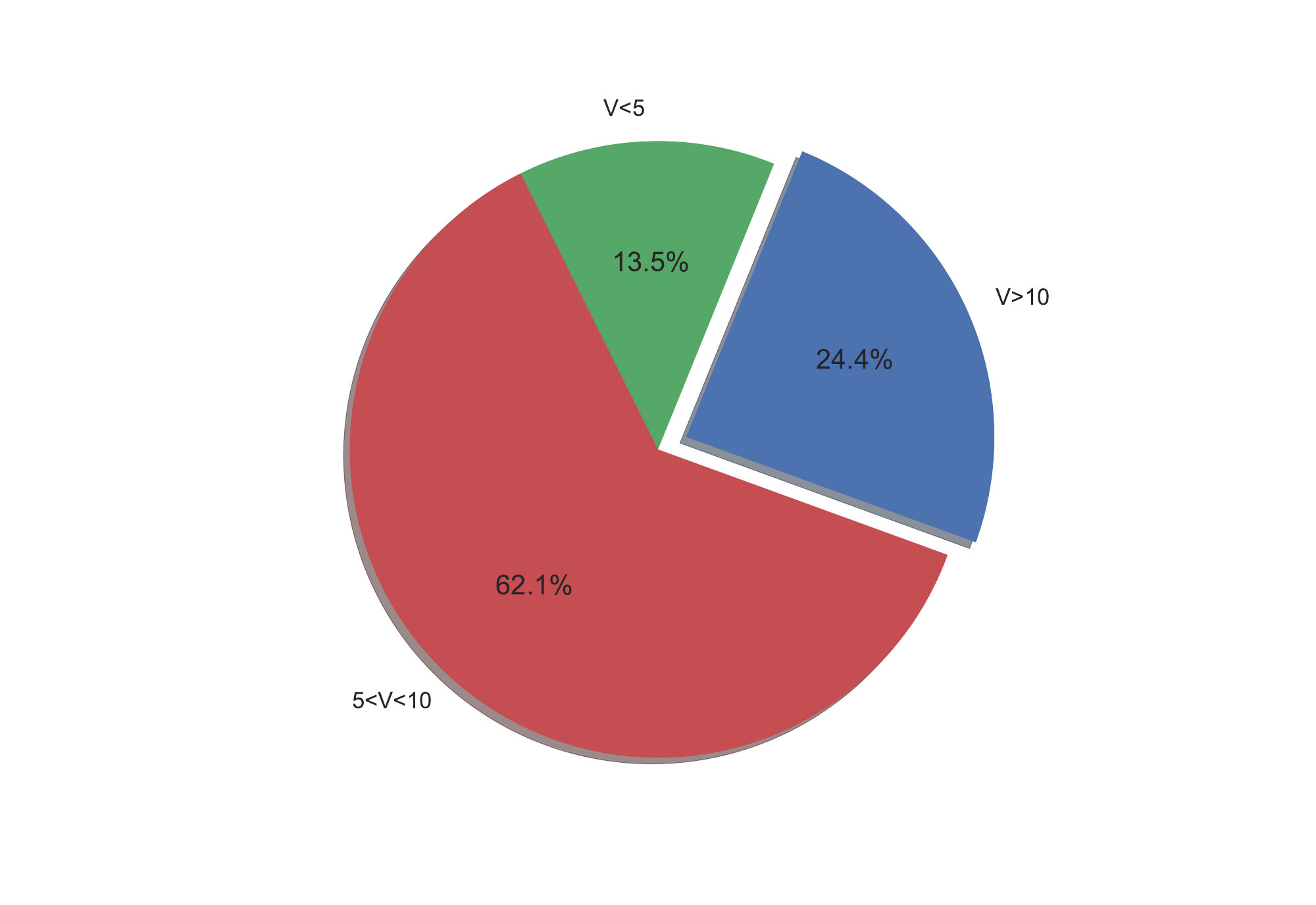}
   \end{tabular}
   \end{center}
   \caption[example] 
   { \label{fig_sample} 
Histogram (left) and pie chart (right) of the magnitudes of the stars used for this analysis. The histogram corresponds to all the near-infrared observations (IRDIS and IRDIFS mode) taken between January 1st 2015 and May 1st 2017 .}
   \end{figure}

The distribution in magnitude for the sample considered here is shown in Fig. \ref{fig_sample}.

\section{Strehl and dependance on atmospheric conditions}
\label{sec_strehl}

\subsection{Strehl and seeing}
\label{sec_strehl_vs_seeing}

\begin{figure} [ht]
 \begin{center}
 \includegraphics[height=8.5cm]{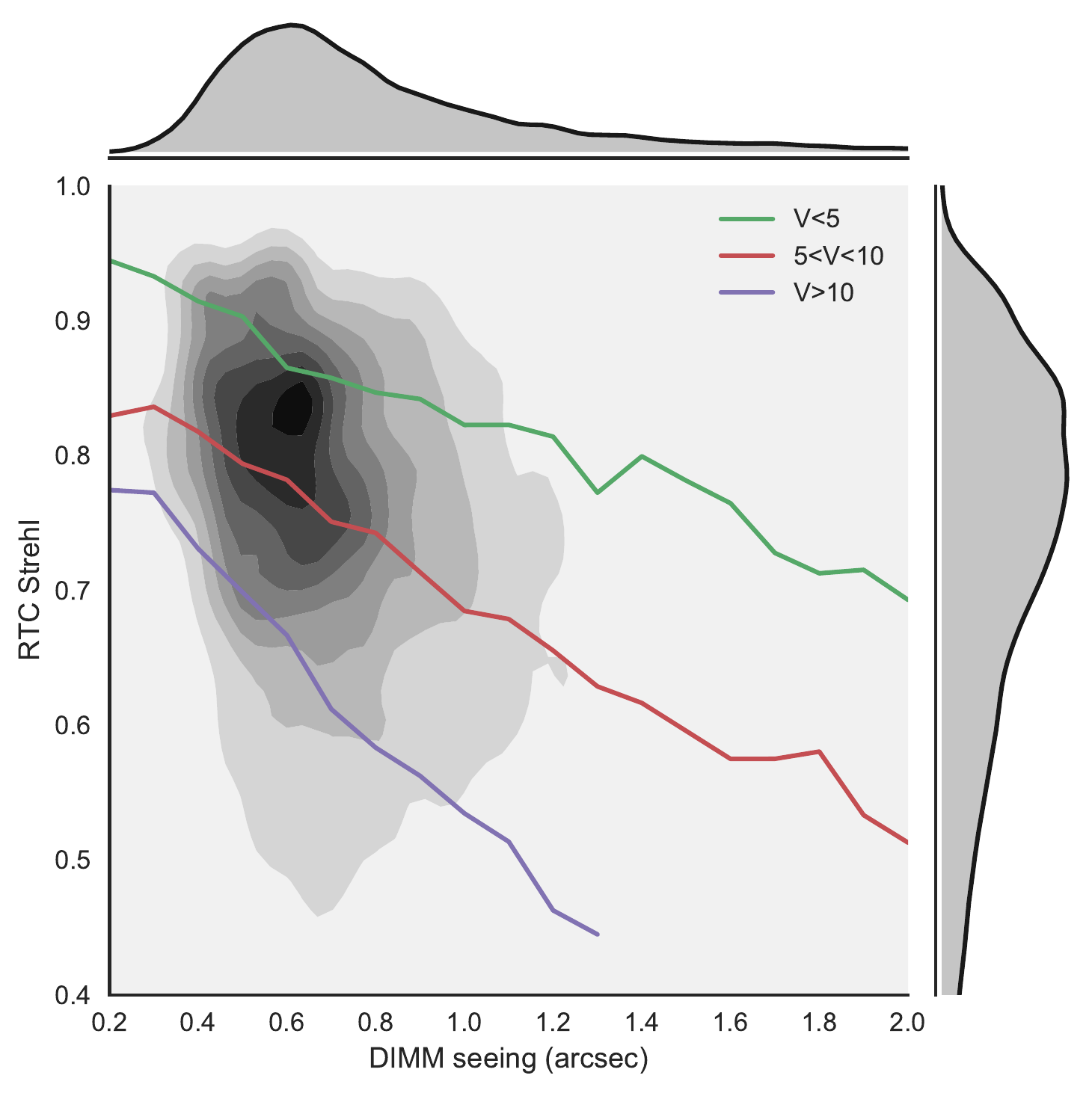}
 \includegraphics[height=8.5cm]{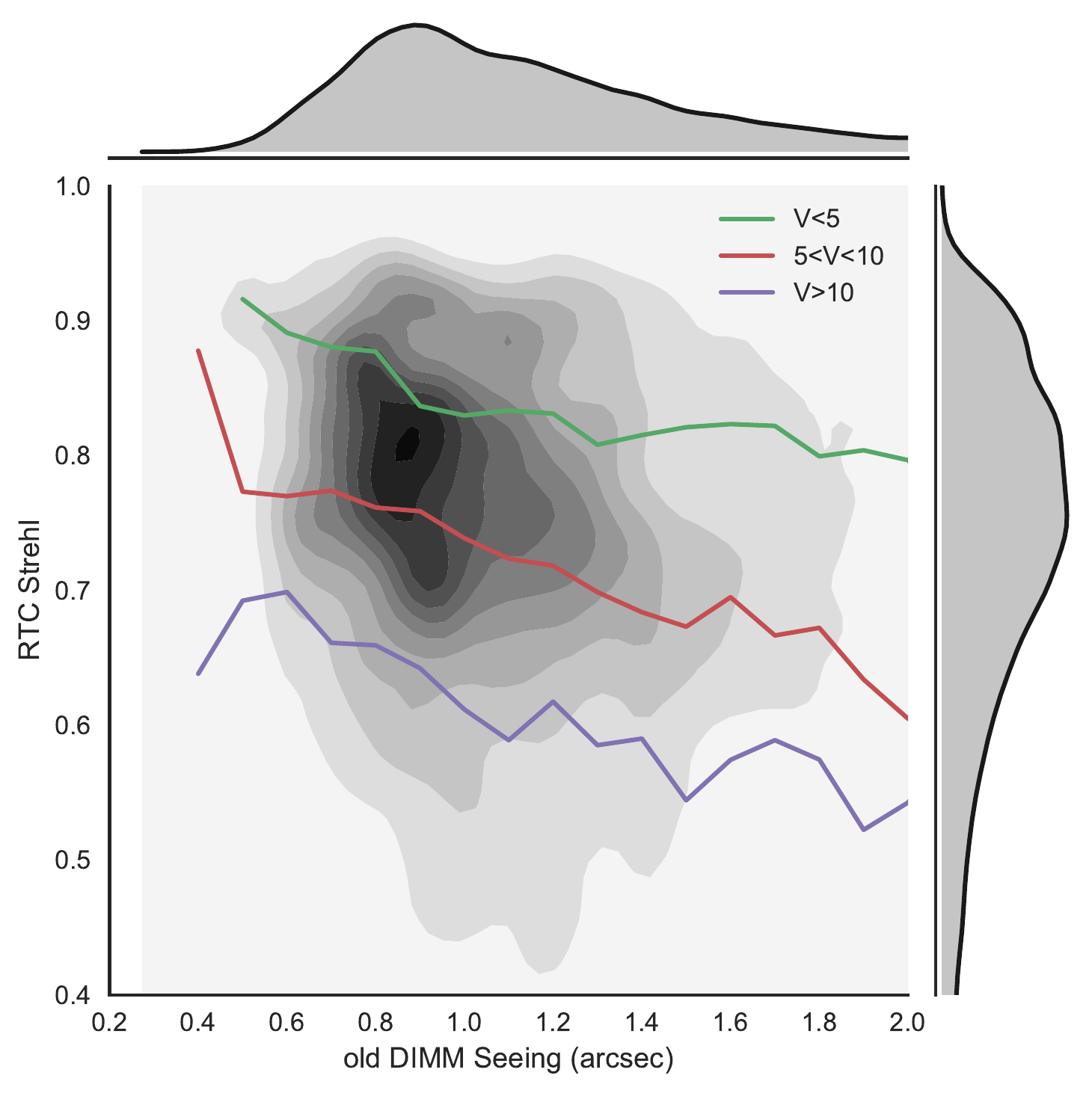}
 \end{center}
 \caption[example] 
{ \label{fig_strehl_vs_seeing}
Distribution of the RTC Strehl as a function of the seeing measured by the new DIMM (left figure) and the old DIMM (right figure). The data points were grouped into three different magnitudes (same colour code as in Fig.\ref{fig_sample} right) and into 0.1\arcsec{} seeing bins to to show how the RTC Strehl depends on the seeing for these different respective magnitudes.
}
\end{figure}

On April 2nd 2016, the measurements of a new Differential Image Motion Monitor (DIMM\cite{Sarazin1990}) became publicly available in the ESO archive. This DIMM is located at a more favourable location on the Paranal platform than the previous DIMM, resulting in more reliable seeing measurements. However, for reference we provide the distribution of the Strehl as a function of both the old and current DIMM in the left and right panels of Fig. \ref{fig_strehl_vs_seeing} respectively. There are 117\,000 telemetry data points with the seeing measurement from the old DIMM and 113\,000 from the current DIMM. 

This plot confirms the good performance level of SAXO\cite{Fusco2016}, with a median Strehl between 80\% and 90\% in good seeing conditions. The large scatter in the grey distributions indicates, as expected, that the seeing is not the only parameter influencing the Strehl (see also section \ref{sec_strehl_vs_tau0} and \ref{sec_strehl_vs_R}). For a user whose science case requests a high Strehl above 80\% or 90\%, specifying the seeing alone as a user constraint for his/her service-mode observations is therefore not an adequate situation. Even with a stringent seeing constraint of 0.6\arcsec, the statistics show that in 25\% of the time, he/she could end up with a Strehl below 75\% for a star of magnitude R between 5 and 10, and below 64\% for a star of magnitude above 10. These quantities correspond to the first quartiles of the Strehl probability distribution for a seeing below 0.6\arcsec, and will be later discussed in the summary plot of Fig. \ref{fig_strehl_vs_seeing_summary_threshold}.

Despite this scatter, the coloured curves, binned in steps of 0.1\arcsec{} seeing, indicate a general trend: a linear decrease in Strehl with the seeing. This decrease is on average 0.9\% Strehl for an increase in seeing of 0.1\arcsec. The slope is slightly steeper for fainter stars, indicating that the Strehl is more sensitive to changes for such objects. 

\begin{figure} [ht]
 \begin{center}
 \includegraphics[height=8cm]{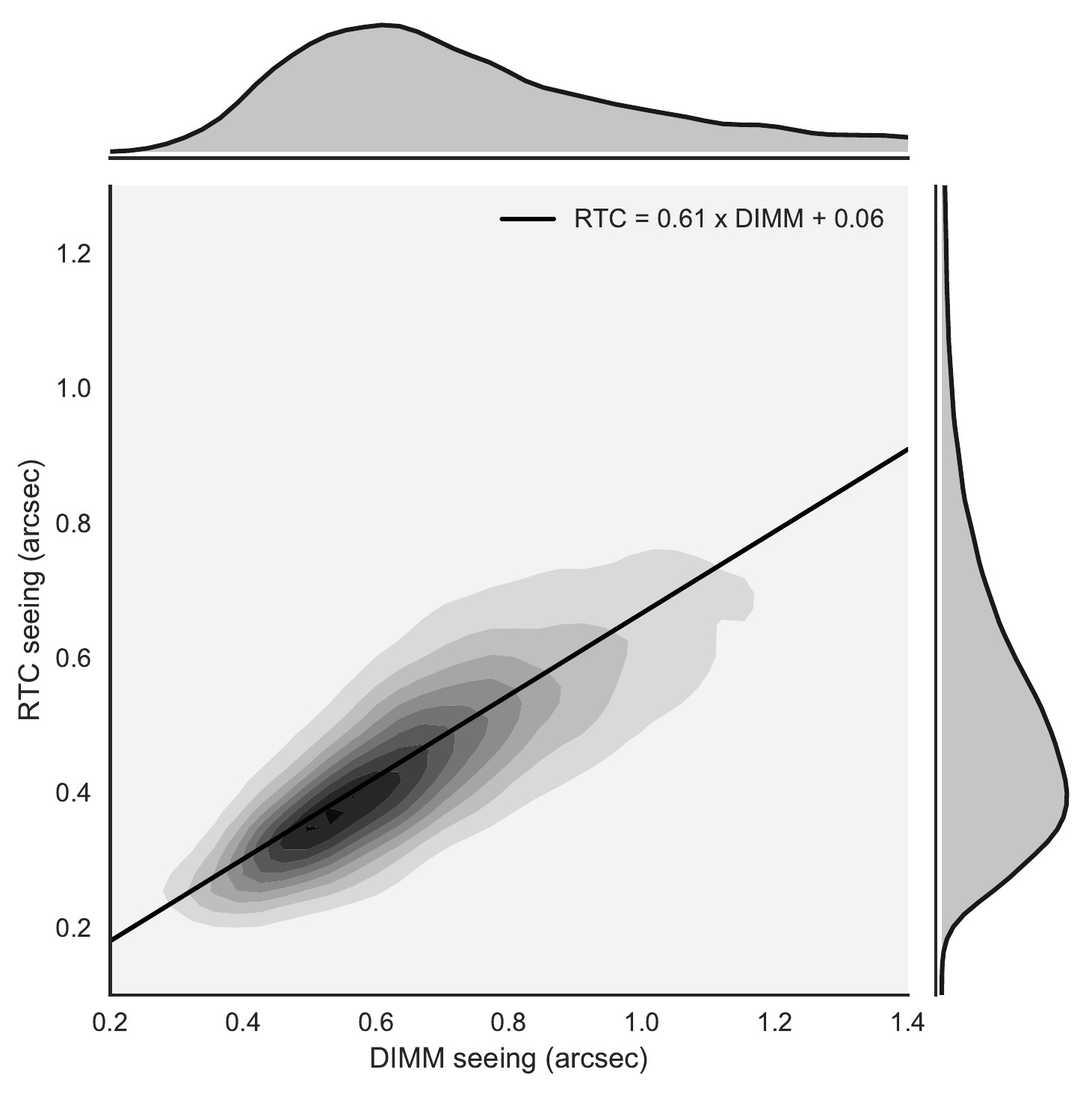}
 \end{center}
 \caption[example] 
{ \label{fig_RTC_vs_DIMM}
Distribution of the RTC seeing as a function of the DIMM seeing. The black curve is a linear fit between 0.35\arcsec{} and 1\arcsec.
}
\end{figure}

As explained in section \ref{sec_scope}, the RTC also provides an estimate of the seeing\footnote{The estimation of the seeing by the RTC is made by reconstructing the open-loop wavefront, projecting it on the Karhunen-Lo\`{e}ve (KL) basis, and computing the temporal autocorrelation of each KL coefficient\cite{Fusco2004} , which depends on the Fried parameter $r_0$ and the turbulence outer scale $L_0$.}. We decided to use the DIMM measurements for the seeing, rather than the RTC estimations for different reasons. First, the accuracy of the RTC estimations depends on the flux received on the WFS. Second, the RTC estimations are always smaller than the DIMM measurements, and we do not have enough open-loop images yet to conclude on the value of the seeing closer to the image quality in the science frames. The difference between the RTC and DIMM measurements for the seeing could come from the difference in the turbulence seen by the DIMM and the telescope, as the DIMM is located 7m above the platform and the telescope is located in a 30m high dome which shields it from the wind. Conversely, this difference may also come from the turbulence outer scale which may influence the RTC estimate and was set to 25m in the RTC algorithm. In any case, the RTC and DIMM seeing show a narrow linear dependance, as visible in Fig. \ref{fig_RTC_vs_DIMM}, therefore the conclusions from Fig. \ref{fig_strehl_vs_seeing} remain valid whether one considers the RTC seeing or the DIMM seeing.

\subsection{Strehl and coherence time}
\label{sec_strehl_vs_tau0}

\begin{figure} [ht]
 \begin{center}
 \includegraphics[height=12cm]{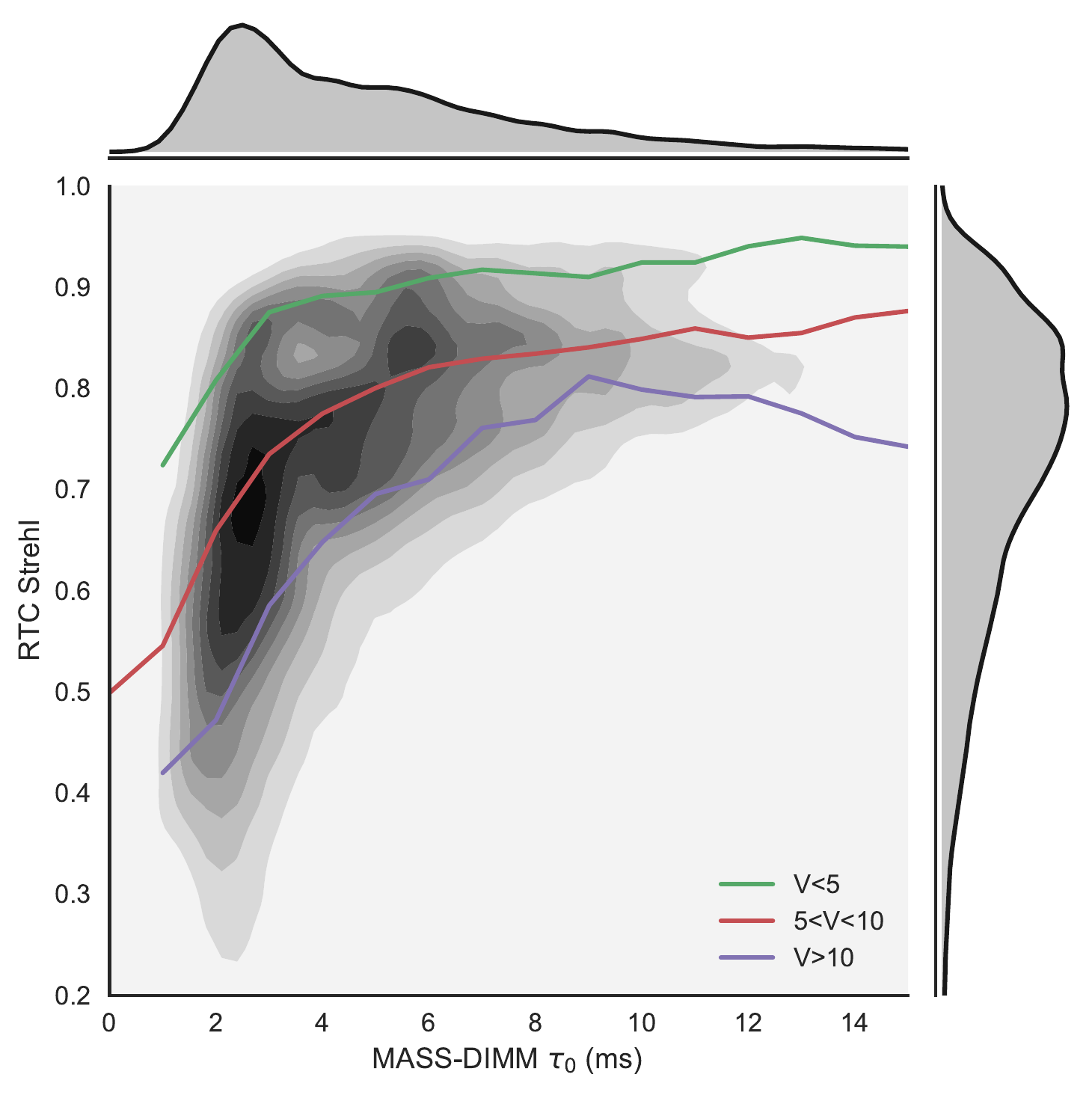}
 \end{center}
 \caption[example] 
{ \label{fig_strehl_vs_tau0}
Distribution of the RTC Strehl as a function of the coherence time as measured by the MASS-DIMM. The three coloured curves show the median RTC Strehl values per 0.1\arcsec{} seeing bin, for the three magnitude classes of target stars as defined in Fig.\ref{fig_sample} (right).
}
\end{figure}

When combined with the MASS\cite{Kornilov2003,Kornilov2007} (Multi-Aperture Scintillation Sensor), the DIMM provides measurements of the coherence time. Among the telemetry data points, 110 000 points have a simultaneous measurement of the coherence time $\tau_0$ from the MASS-DIMM. Fig. \ref{fig_strehl_vs_tau0} shows the distribution of the Strehl as a function of the coherence time for this sample.

The impact is large with a steep rise in Strehl with the coherence time, before a shallower increase or even a plateau for larger coherence times. The transition between these two regimes depends on the target magnitude: it occurs at 3ms for bright stars ($V\leq5$, green curve in Fig. \ref{fig_strehl_vs_tau0}), at 4ms for stars of intermediate brightness ($5 \leq V \leq 10$, red curve) and only at 7ms for faint stars ($V\geq10$, purple curve). 

This shows that for low coherence times, the WFS is limited by the temporal bandwidth error.
This is in agreement with the laboratory and first on-sky measurements from the SPHERE RTC, described in details in Petit et al. 2014\cite{Petit2014} who derived a temporal bandwidth of 70Hz with an initial frame rate of 1.2 kHz. This bandwidth corresponds to a time scale of 14ms, and it comes as no surprise that the performance degrades when the turbulence time scale $\tau_0$ is below 5\,ms. 

\subsection{Strehl and star magnitude}
\label{sec_strehl_vs_R}

The SPHERE instrument can close the AO loop on targets up to an R magnitude of 15 in degraded conditions. Xu et al. (2015)\cite{Xu2015} describe the Strehl and contrast performance obtained on a faint white dwarf of magnitude R=14.2. This capability to maintain decent performance for faint targets is a real asset of the SPHERE instrument compared to its extreme-AO counterparts such as the Gemini Planet Finder (GPI\cite{Macintosh2014}) or the Subaru Coronagraphic Extreme AO (SCExAO\cite{Guyon2010}). Over the complete sample of stars observed by SPHERE during the first two and a half years of operations, 103 stars are fainter than magnitude R=12.

\begin{figure} [ht]
 \begin{center}
 \includegraphics[height=12cm]{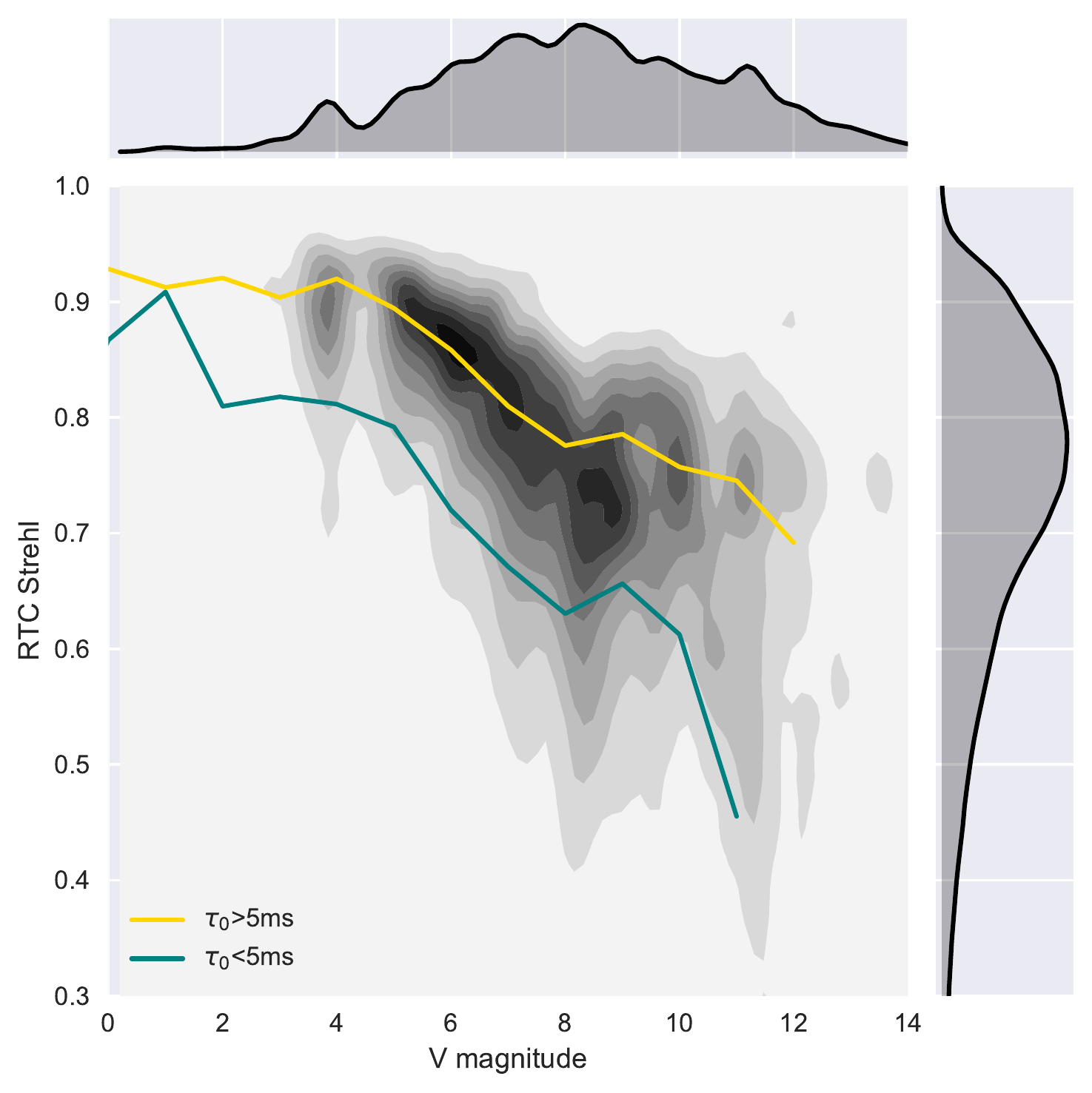}
 \end{center}
 \caption[example] 
{ \label{fig_strehl_vs_Rmag}
Distribution of the RTC Strehl as a function of the V magnitude of the target star. The two coloured curves show the median RTC Strehl values per bin of one magnitude, for the two classes of conditions: low coherence time (green) and long coherence time (yellow).
}
\end{figure}

Fig. \ref{fig_strehl_vs_Rmag} shows the distribution of the RTC Strehl as a function of the star magnitude in V (more readily available than the R magnitude). 
It shows a decrease in the Strehl beyond mag 6. There seems to be a slight inflection for a magnitude of 8, followed by a further decrease beyond magnitude 9.
We made a distinction in Fig.\ref{fig_strehl_vs_Rmag} between low (green curve) and high coherence times (yellow curve). The general trend is similar for both conditions up to magnitude 10, with an offset between the two curves that comes as no surprise given the results shown in section \ref{sec_strehl_vs_tau0}. Beyond magnitude 10, the decrease in Strehl is very steep for fast-evolving turbulence conditions, and smoother for a slower turbulence regime. The curves were stopped after magnitude 11 and 12 respectively, as few data points are available and the RTC Strehl measurement becomes unreliable in the photon starving regime. 

The decrease after magnitude 6 is unexpected from a design point of view (see Fusco et al. 2006 \cite{Fusco2006}) as the WFS noise is not expected to be the dominant error contributor below R=9 \cite{Fusco2016}. Possible effects currently under investigation are a bias in the Strehl estimation by the RTC for stars between magnitude 5 and 9 (see also next section for the validation), or effects coming from the different regimes of the AO loop (AO frequency, WFS gain, etc.) that are set as a function of the R magnitude.

\subsection{Summary}
\label{sec_summary}

 \begin{figure} [ht]
 \begin{center}
 \includegraphics[height=20cm]{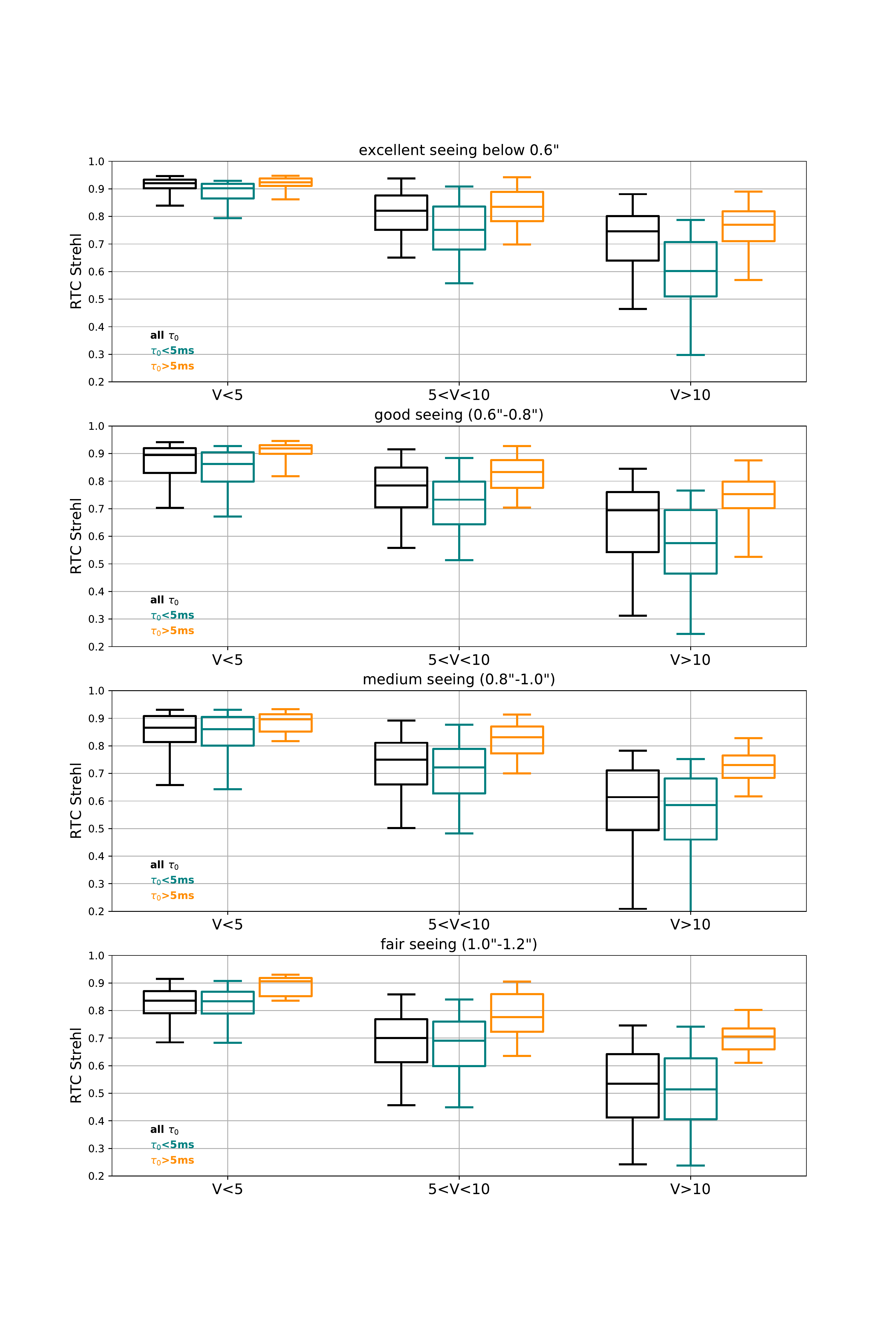}
 \end{center}
 \caption[example] 
{ \label{fig_strehl_vs_seeing_summary_threshold}
Summary of the dispersion of the RTC Strehl in different seeing conditions (from top to bottom: below 0.6\arcsec, from 0.6\arcsec{} to 0.8\arcsec, from 0.8\arcsec{} to 1.0\arcsec and from 1.0\arcsec{} to 1.2\arcsec). The boxes indicate the 25\%, 50\% and 75\% quartiles of the distribution, while the whiskers indicate the 5 and 95\% percentiles. In each plot, we grouped the data by star brightness and by coherence time, as measured by the MASS-DIMM.
}
\end{figure} 

A summary of the distribution of the Strehl as measured by the RTC is provided in Fig. \ref{fig_strehl_vs_seeing_summary_threshold}. Currently the seeing is still the only parameter that SPHERE service mode users can set for their observations. In this respect, Fig. \ref{fig_strehl_vs_seeing_summary_threshold} shows the range of Strehl that can be expected from the instrument, given the seeing constraint (0.6\arcsec, 0.8\arcsec,1.0\arcsec or 1.2\arcsec), depending on the target magnitude and the coherence time (currently not a constraint). 
It shows that the Strehl distribution changes significantly between low and high coherence times (green and orange box plots). This conclusion is especially striking for faint stars of magnitude larger than 10. In this case, the third quartile of the Strehl distribution with a fast coherence time (green box plots) corresponds to the first quartile of the distribution with a long coherence time (orange curve). The width of the distribution is twice as large in the two cases. As an example, even in very good seeing conditions below 0.6\arcsec{}, 5\% of the faintest targets are observed with a Strehl below 30\% in the case of a low coherence time. However in case of long coherence time, the 5\% percentile on faint star rises to 56\% (top right box plot in Fig. \ref{fig_strehl_vs_seeing_summary_threshold}).

This suggests that coherence time should be included as a constraint for service mode observations, in order to be able to allocate the nights with longer coherence times to the faintest targets. Here, service mode users requesting faint target stars could ask for longer coherence times. The observatory could then guarantee a much narrower Strehl distribution (orange box plots in Fig. \ref{fig_strehl_vs_seeing_summary_threshold}) compared with no constraint on the coherence time (black box plots).


\section{Contrast}
\label{sec_contrast}

\begin{figure} [ht]
 \begin{center}
 \includegraphics[height=6cm]{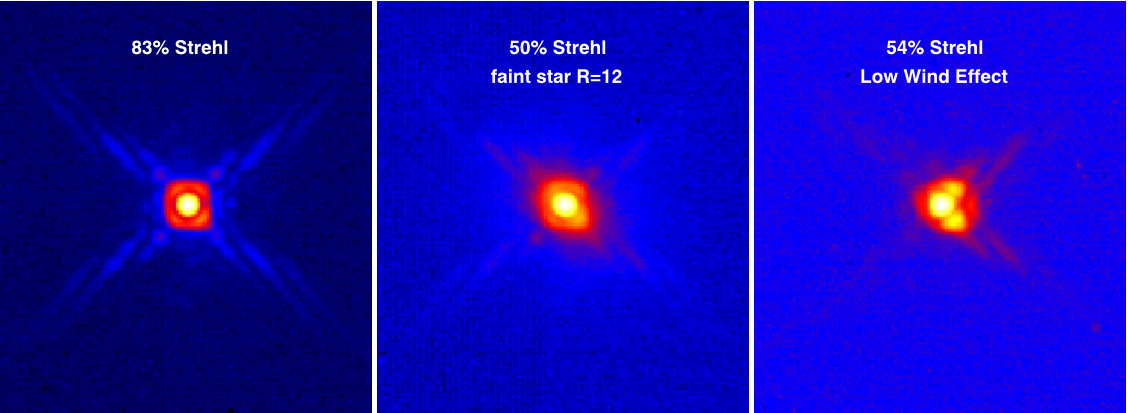}
 \end{center}
 \caption[example] 
{ \label{fig_PSF_examples}
Examples of a nominal PSF (left) and two PSFs affected by a degradation in Strehl due to the faintness of the star (middle) and the low wind effect (right) creating two secondary lobes, hence their nickname Mickey Mouse ears. These PSFs correspond to the VLT pupil, combined with an apodizer and an undersized Lyot stop which covers the spiders and the DM bad actuators. 
}
\end{figure}

\begin{figure} [ht]
 \begin{center}
 \includegraphics[height=5.5cm]{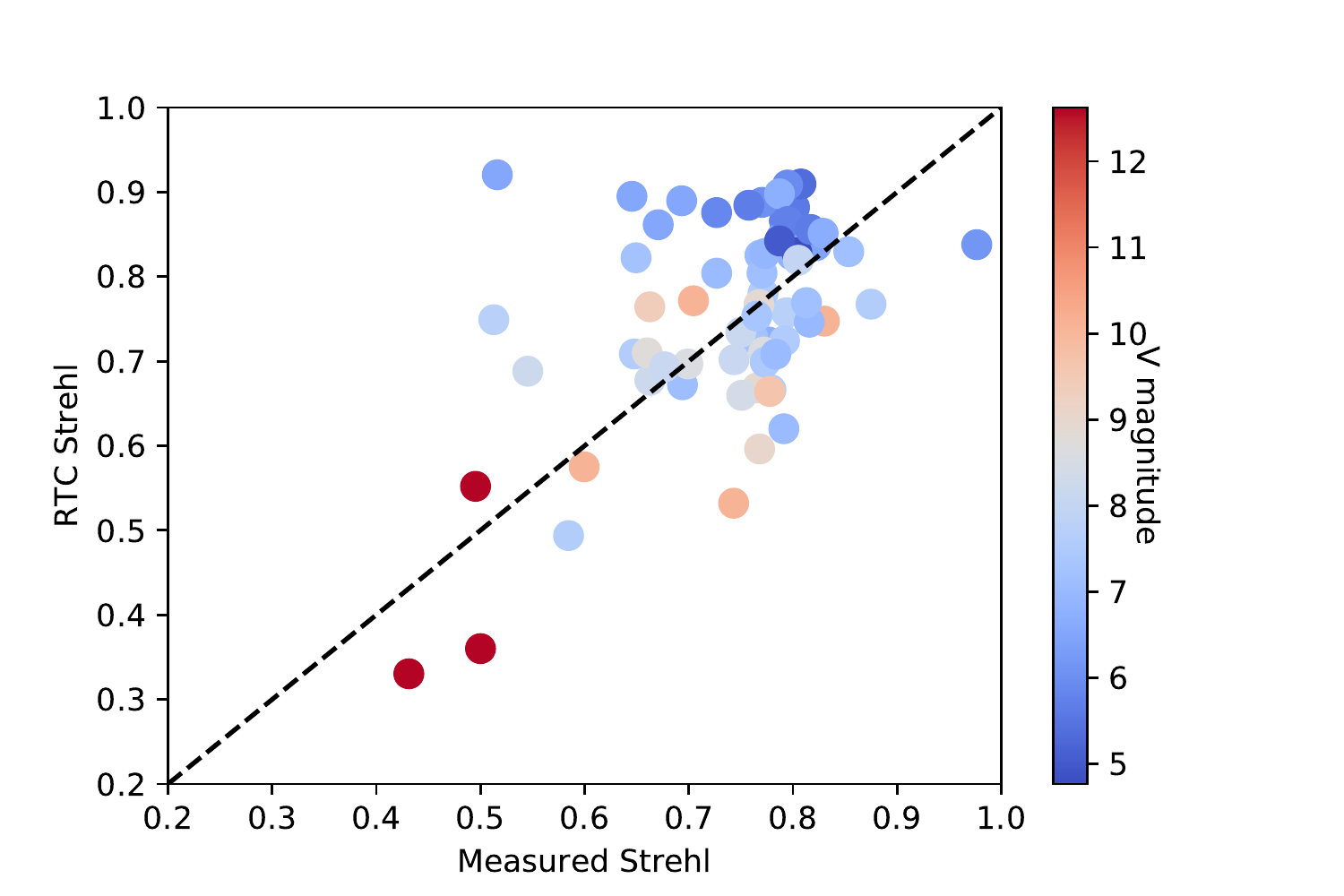}
 \includegraphics[height=5.5cm]{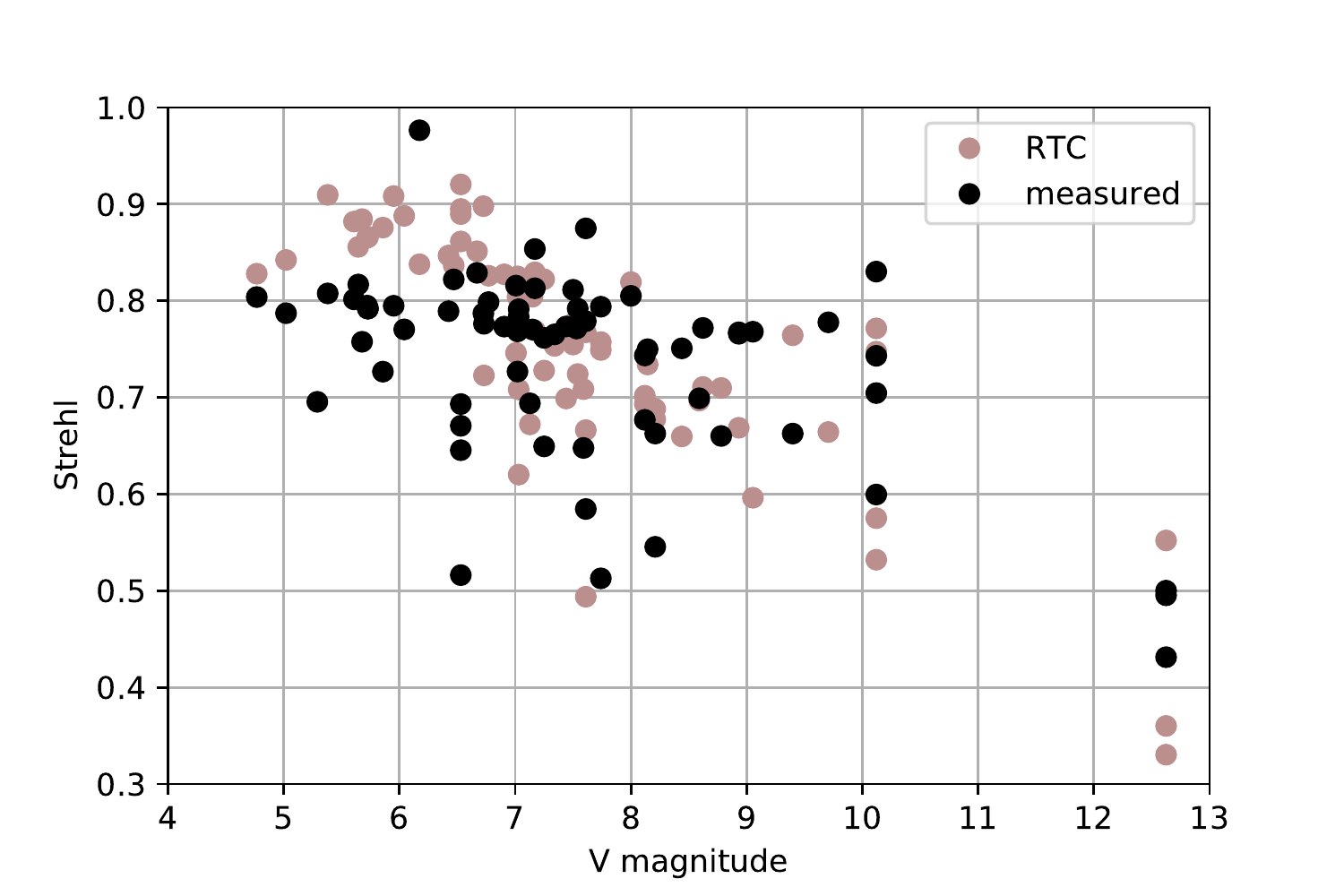}
 \end{center}
 \caption[example] 
{ \label{fig_strehl_validation}
Left: RTC Strehl vs. measured Strehl for a reduced sample. The colour scale indicates the star brightness in V. Right: Measured and estimated Strehl vs. magnitude for the same reduced sample.
}
\end{figure}

The ultimate parameter SPHERE users are interested in is not the Strehl but the contrast. So far we used the Strehl ratio as a convenient proxy because it is estimated by the AO system and is easily available without requiring data processing. To understand the relationship between contrast and Strehl, we used a reduced sample of 55 different stars observed as part of the SPHERE High-Angular Resolution Debris Disc Survey (SHARDDS\footnote{ESO program ID 096.C-0388(A) and 097.C-0394(A), with a total allocation time of 55h on VLT/SPHERE (Milli et al. in prep.)}, PI: J. Milli), in the broad-band H filter(centred at 1.625\micron, width 291nm) with the apodized Lyot coronagraph of diameter 185mas. The SHARDDS program is an open-time program on SPHERE to search for new debris discs in scattered light, around young nearby stars which have a large infrared excess but no disc detection at the date of the ESO P96 call for proposal (2015). It has already led to several disc detections\cite{Wahhaj2016,Choquet2017} . The targeted stars range from magnitude 5 to 10, with the exception of one star of magnitude V=12. The Strehl was first measured in non-coronagraphic exposures\footnote{The Strehl measurement is done by computing the optical transfer function of the measured non-coronagraphic image, and comparing it to the theoretical optical transfer function including all the optical elements of the system (apodiser, Lyot stop...)}, to validate the RTC estimation. A seeing constraint of 1\arcsec{} was set for the observations of this program. This constraint was not met for all observations (and the observations were repeated in this case), however we included all observations for our analysis , even those with a seeing value greater than 1\arcsec.
The comparison between the measured Strehl and the RTC-estimated Strehl is provided in Fig. \ref{fig_strehl_validation} (left). Despite a few outliers, there is a good agreement between the RTC estimations and the measurements. After investigating the outliers, we could relate most of the overestimations of the Strehl by the RTC to two issues: 
\begin{itemize}
\item the presence of the low-wind effect\cite{Sauvage2016} . This affects about 20\% of the SPHERE observations and is unseen by the wavefront sensor, hence it is not reflected in the RTC Strehl estimation. An illustration of a Point-Spread-Function (PSF) affected by this problem is given in Fig. \ref{fig_PSF_examples} (right).
\item the bad calibration of the reference slopes of the WFS during two weeks in January 2016, which affected a few points in this subsample.
\end{itemize}

This reduced subsample also allowed us to investigate the unexpected decrease of the Strehl as a function of the target magnitude, which occurs beyond magnitude 6 instead of the expected value of magnitude 9 to 10 (see section \ref{sec_strehl_vs_R}). Fig. \ref{fig_strehl_validation} shows the measured  and RTC-estimated Strehl as a function of the V magnitude of the target (brown and black points respectively). Although the same decrease after magnitude 6 to 7 is visible in the estimated Strehl scatter plot, this trend does not appear for the measured Strehl. This supports the assumption of the bias in the RTC measurement, presented in section \ref{sec_strehl_vs_R}, however additional data are currently being analysed to confirm this. 

\begin{figure} [ht]
 \begin{center}
 \includegraphics[height=10cm]{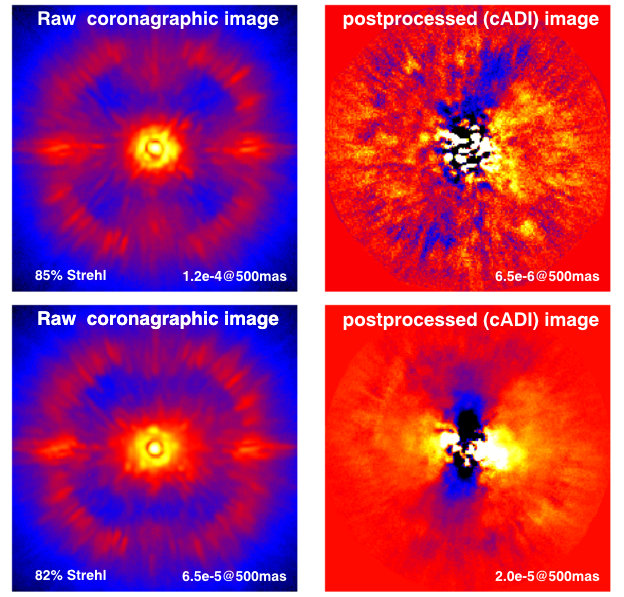}
 \end{center}
 \caption[example] 
{ \label{fig_coro_images_examples}
Examples of raw coronagraphic images of a few seconds exposures (left) and after applying classical ADI (right) on a 40min sequence of pupil-stabilised observations. The first raw image illustrates the case of nominal conditions under a good seeing and coherence time, while the second raw image illustrates the case of a low coherence time due to high-altitude wind. Although the PSF elongation due to the wind is not clearly visible in the raw coronagraphic image because it is hidden in the speckle halo, the post-processed image reveals the typical butterfly pattern in the direction of the wind (here East-West, the predominant jet stream direction above Paranal). The colour scale is linear and the stretch is 100\% of the pixel brightness range for the left images, and 98\% for the right images. The image is 2.45\arcsec{} on the side.
}
\end{figure}

 We computed the raw contrast as follows: the frames for each observation were sky-subtracted, flat-fielded and bad pixel-corrected using the official SPHERE Data Reduction and Handling pipeline\cite{Pavlov2008} in order to make a temporal cube of frames. Each coronagraphic frame was re-centred using the set of four satellite spots imprinted in the image during the centring sequence. This sequence is obtained by applying a waffle pattern to the deformable mirror and was done prior and after the deep science coronagraphic observations. A few frames were discarded based on a flux criteria within an annulus centred outside the edge of the coronagraph. An example of coronagraphic frames is shown in Fig. \ref{fig_coro_images_examples} left. We applied the same cosmetic reduction to the images obtained out of the coronagraphic mask, with a neutral density, which were used for the calibration of the contrast with respect to the star, and for the direct Strehl measurement. Three of those frames are presented as an illustration in Fig. \ref{fig_PSF_examples}, in a nominal case (left image) and in two degraded Strehl conditions (middle and right image).
The contrast was then computed using the Vortex Imaging Pipeline\cite{Gomez2017} that computes the integrated flux in apertures with a size of one resolution element placed at all azimuths and at increasing radii around the star center. For a given radius, the $5\sigma$ contrast is expressed as five times the standard deviation of those integrated fluxes, corrected by the small-sample statistics \cite{Mawet2015} . Two types of contrast are shown in this analysis:
\begin{itemize}
\item the raw contrast, computed on the median coronagraphic image of each observation (Fig. \ref{fig_coro_images_examples} left).
\item  the contrast after applying classical Angular Differential Imaging (ADI\cite{Marois2006}) on the data cube (Fig. \ref{fig_coro_images_examples} right). In this latter case, the contrast is corrected for the throughput of the algorithm by injecting fake companions to calculate this radius-dependent throughput.
\end{itemize}

\begin{figure} [ht]
 \begin{center}
 \includegraphics[height=5.5cm]{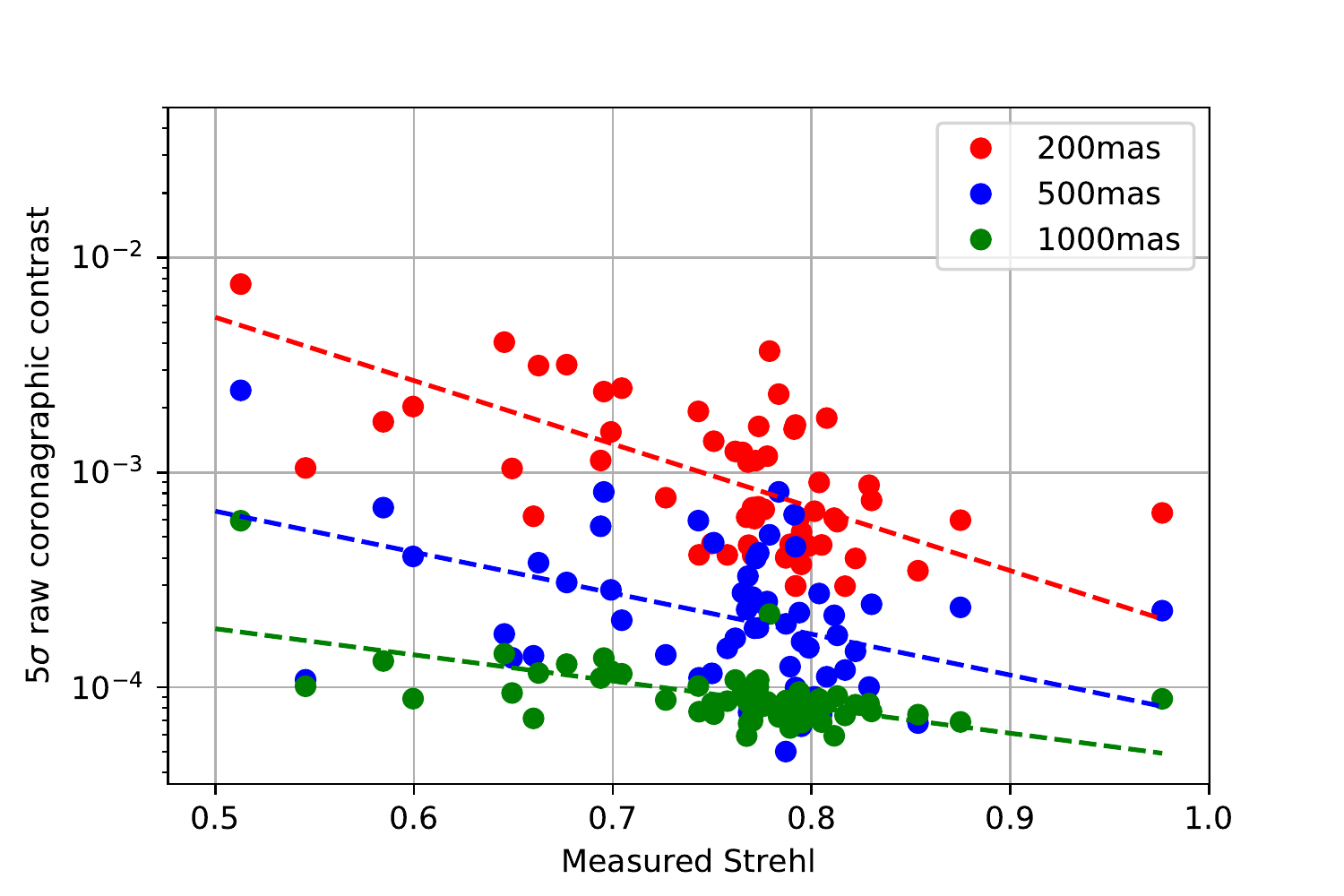}
 \includegraphics[height=5.5cm]{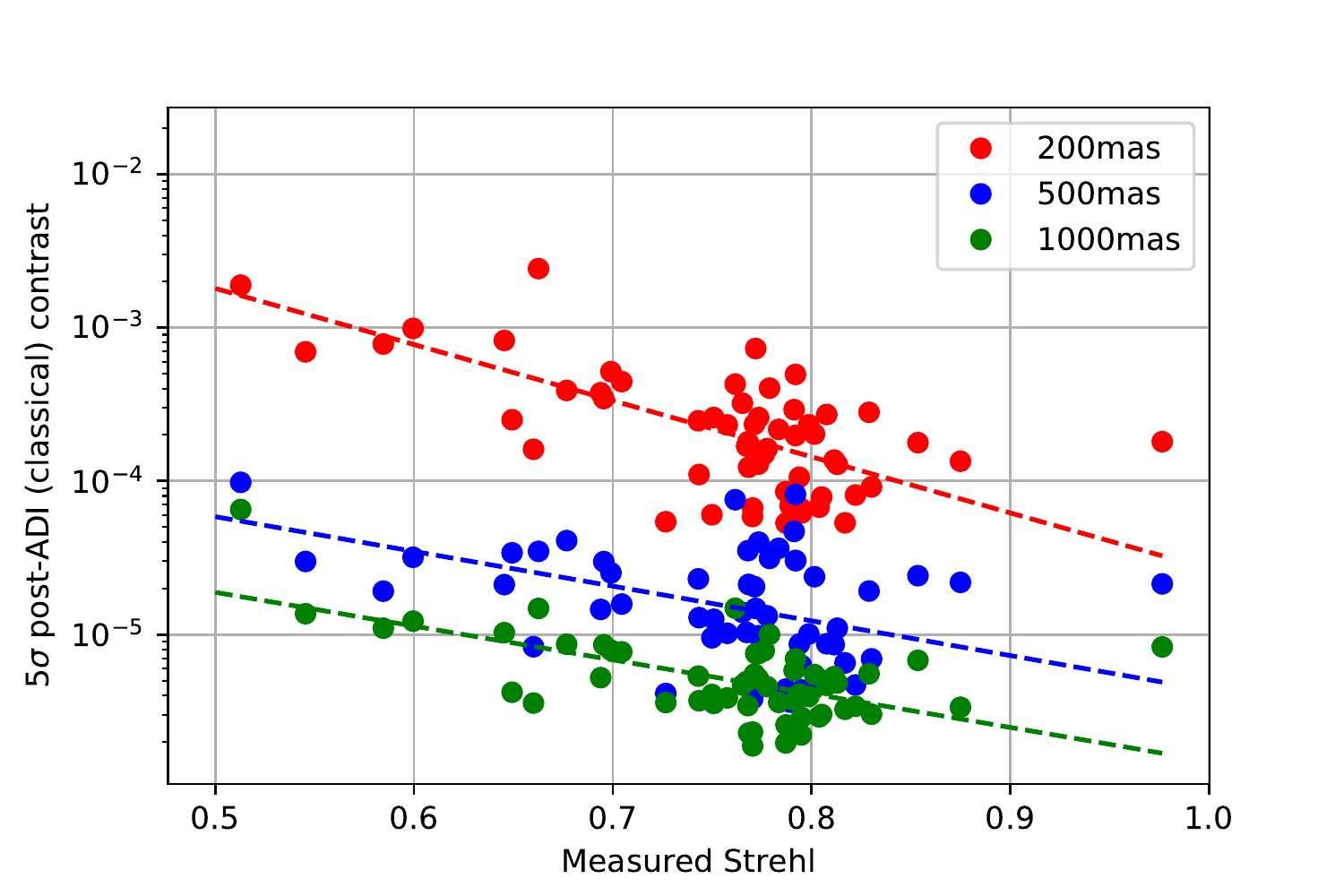}
 \end{center}
 \caption[example] 
{ \label{fig_contrast_vs_strehl}
Left: $5\sigma$ raw coronagraphic contrast at 3 different separations as a function of the measured Strehl. Right: throughput-corrected $5\sigma$ contrast after applying classical ADI as a function of the measured Strehl for effective observations of 40min around meridian.
}
\end{figure}

The raw coronagraphic contrast (Fig. \ref{fig_contrast_vs_strehl} left) at 500 or 200mas shows a dependence on the Strehl, as expected from theoretical considerations\cite{Serabyn2007} . These two separations are indeed well within the AO control radius of about 840mas in the H-band. This dependence remains clearly visible on post-processed data after applying ADI (Fig. \ref{fig_contrast_vs_strehl} right), especially at 200mas, despite other factors also coming into considerations such as the field rotation and the stability of the conditions. At this separation, the scatter plot shows that one can easily gain one order of magnitude in contrast by increasing the Strehl by 10\%. 
Outside the AO correction radius, at 1000mas, there is little dependance of the contrast on the Strehl, except for a few outliers taken in bad conditions.

As part of this study, the dependance of the post-ADI contrast on parameters tracing the stability of the conditions was also investigated. We considered the dispersion in the seeing, coherence time or equivalent velocity, as well as the dispersion in Strehl, during the duration of the pupil-stabilised sequence. No significant correlation could be drawn from this limited sample and a larger study including more than 55 observations is required to investigate these dependencies statistically. 

We also highlight that one clear cause of contrast degradation in the post-ADI contrast is the presence of a smooth halo within the AO-corrected region in the direction of the wind, when a high wind was present in the data (most of the time a high-altitude wind related to the jet stream at 200mbar). This halo is rotating in the pupil-stabilised data set because it is fixed on the sky. In post-ADI frames, it therefore appears as a brighter elongation along the wind direction, with negative counterparts at $90^\circ$. We illustrate this effect in Fig. \ref{fig_coro_images_examples} (bottom right). Although the effect is not visible in the raw coronagraphic image (bottom left) because it is below the floor of speckle noise, it appears after subtraction of the static part of the PSF as done in classical ADI. In this example, although the $5\sigma$ raw coronagraphic contrast is very good (below $10^{-4}$ at 500mas), the gain through star-subtraction is only by a mere factor 3 whereas post-processing through classical ADI typically improves the contrast by a factor 10 to 50. This problem can be addressed in different ways. In hardware, increasing the temporal bandwidth of SAXO is one solution to mitigate this effect. This is considered in a forthcoming upgrade of the instrument. In post-processing, PSF reconstruction algorithms or phase retrieval techniques\cite{Ygouf2013} are currently under investigation.  

\section{Conclusions and perspectives}
\label{sec_conclusions}

In the first two and a half years in which SPHERE has been offered to the community, this instrument has observed more than 1000 stars from magnitude -2 to 15, delivering high-Strehl, high-contrast images, as well as a wealth of AO telemetry data that were analysed in this study. We confirmed that the performance level reached during these regular operations is aligned with the first on-sky results gathered during the commissioning runs\cite{Fusco2016} . We presented the effect of the seeing, coherence time and star brightness on the AO performance. The fitting error is already well established, with this effect clearly visible in the non-corrected region of the PSF beyond 800mas in the images presented here. This study reveals however the significant impact of the temporal bandwidth error for SPHERE, given the distribution of coherence times existing at the Paranal observatory. The effect of low coherence times on the Strehl was clearly demonstrated in a statistical way, impacting directly the corrected region of the coronagraphic images, even under good seeing conditions. A similar conclusion was also reached concerning the high-contrast instrument GPI\cite{Bailey2016} . The dependence on stellar magnitude was also shown, indicating optimal AO performance up to magnitude 6. A more detailed investigation is required to understand if the decrease in Strehl beyond magnitude 6 is real and what causes it. The contrast was analysed for a subsample of 55 stars, indicating very good $5\sigma$ raw contrast performance of $10^{-4}$ to $10^{-5}$ at 500mas, correlated with the Strehl. This analysis now requires a larger sample to be able to explain the dependence of the contrast reached in raw and post-ADI frames on the system state, atmospheric parameters and AO star properties.  Further statistical methods are being looked at to be able to do accurate performance predictions and introduce new user constraints for a higher efficiency and science yield of this community instrument.


\acknowledgments 
 
This research made use of Astropy, a community-developed core Python package for Astronomy (Astropy Collaboration, 2013) and of  the SIMBAD database, operated at CDS, Strasbourg, France. It also used the search engine Elasticsearch, a distributed search and analytics engine, in combination with Kibana for data visualisation.  Those tools are part of a technical database installed at the Paranal Observatory. 

\bibliography{biblio} 
\tiny
\bibliographystyle{spiebib} 

\appendix    

%
%

\end{document}